# Sewer Rats in Teaching Action

## An explorative field study on students' perception of a game-based learning app in graduate engineering education

Heinrich Söbke[1] and Maria Reichelt[2]

[1] Bauhaus-Universität Weimar, Bauhaus-Institute for Infrastructure Solutions (b.is), Weimar, Germany
[2] Fachhochschule Erfurt, University of Applied Sciences, Centre for Quality, Erfurt, Germany

```
heinrich.soebke@uni-weimar.de
maria.reichelt@fh-erfurt.de
```

**Abstract.** Game-based technologies and mobile learning aids open up many opportunities for learners; however, evidence-based decisions on their appropriate use are necessary. This explorative study (N = 100) examines the role of game elements in university education using a game-based learning app for mobile devices. The educational goal of the app is to support students in the field of engineering to memorize factual knowledge. The study investigates how the game-based app affects learners' motivation. It analyses the perceived impact and appeal as well as the game elements as an incentive in learners' perception. To realize this aim, the study combines structured methods like questionnaires with semi-structured methods like thinking aloud, game diaries, and interviews. The results indicate that flexible temporal and spatial use of the app was an important factor of learners' motivation. The app allowed more spontaneous involvement with the subject matter and the learners took advantage of an improved attitude toward the subject matter. However, only a low impact on intrinsic motivation could be observed. We discuss reasons and present practical implications.

## 1 Introduction

Using game-based elements in non-gaming contexts like learning settings is an ongoing trend in the process of digitalizing educational fields (e.g., Roppelt 2014). Innovative educational technologies offer the opportunity to arrange new educational settings. According to the motivational-based ARCS model from Keller (2010), learners' attention and interest can be captured through new learning aids and technologies. Approaches to develop game-based learning aids were mainly explored in school settings

(e.g., Karakus, Inal, and Cagiltay 2008), while formal settings of higher education have so far been widely neglected. Ratan and Ritterfeld (2009) found that only 16% of games target age groups including and above college level. However, game elements are in fact used in higher educational settings (e.g., Ebner and Holzinger 2007; Roppelt 2014) as a way to influence learners' motivation and heighten interest in the learning content. Although game-based technologies and mobile learning tools can open up great potential for learning and teaching, evidence-based decisions on their appropriate use are necessary. Therefore, our study investigates the effect of a game-based learning app on learners' motivation. It analyzes the perceived impact and appeal as well as the game elements as one incentive of learners' perception. We analyze these questions using a digital game application—the KanalrattenShooter (KRS) app. This iOS app uses a game-based approach to provide factual knowledge about the discipline of urban wastewater management (Söbke, Chan, Buttlar, Große-Wortmann, & Londong, 2014). The following section considers the KRS app from the perspectives of both serious games and mobile learning.

## 2   Theoretical Framework

### 2.1   First Perspective: Serious Games

Successful commercial video games employ principles of good instruction (Gee 2008), and they are known as learning machines (Gee 2004). In consequence, they constitute versatile instruments in the field of game-based learning (Meier and Seufert 2003; Squire 2011). An essential characteristic of video games is their impact on players' motivation. Well-designed games foster intrinsic motivation, i.e. they are played without having any intentions outside of the game (Ryan and Deci 2000). Thus players can reach a state of complete immersion in the game, called *flow* (Csikszentmihalyi 1990). The capability to evoke intrinsic motivation is a characteristic of well-designed serious games, i.e. games with an additional purpose (e.g. learning) beyond entertainment (Djaouti, Alvarez, and Jessel 2011; Michael and Chen 2005). If there is no interest in playing, games just become "another assignment" (Rockwell and Kee 2011).

Much work has been done to identify and categorize motivational elements as well as player types and learner types. Malone and Lepper (1987) present a taxonomy of intrinsic motivation in the context of learning. Garris, Ahlers, and Driskell (2002) discuss known motivation models and emphasize the relevance of motivation in the context of learning. A key concept in the game context is that specific player types prefer specific motivational elements. Among the first to describe player types has been Bartle (1996), who has identified four playing styles called Achievers, Explorers, Socializers, and Killers in multi-user dungeons (MUDs). Yee (2006) lists motivation subcomponents and relates them to player profiles. Konert, Göbel, and Steinmetz (2013) have tried to link personality profiles to learner types and player types in order to define a foundation for adaptive game-based learning. They conclude that player type and learner type have to be determined individually. Successful learning games must integrate learning content and game mechanics in a way that does not diminish intrinsic motivation (Garris et al. 2002), a requirement that Habgood and Ainsworth (2011) call "intrinsic integration." The challenge of a balanced integration is demanding as a long list of failed attempts suggests (Bruckman 1999; Egenfeldt-Nielsen 2005; Papert 1998).

A reaction to this challenge is not to integrate games and "serious" content, but instead to add game design elements to real-world applications in order to impact intrinsic motivation (Deterding, Dixon, Khaled, and Nacke 2011). An example is a "gamified" learning platform that rewards learning actions and social interaction with points and badges (Kapp 2012; Simões, Redondo, and Vilas 2012).

Besides motivation there are further dimensions to consider in the field of game-based learning. Plass, Homer, and Kinzer (2015) present a field-organizing framework that adds the aspects of *Affect*, *Cognition*, and *Social/Cultural*, illustrating the complexity of game-based learning. As a potential alleviation for this problem we have argued for using established, commercial off-the-shelf games (Söbke, Bröker, & Kornadt, 2013) or proven game mechanics (Söbke, 2015; Söbke et al., 2014). Especially quizzing, seems to be a generally intriguing form of play. This is relevant in our case as KRS is a quiz app in the broader sense.

**2.2  Second Perspective: Mobile Learning Aids**

The research area of Instructional Design (ID) researches design principles of effective learning environments. It is a field of educational technology concerned with the systematic design of learning environments based on empirical evidence and theory (e.g., Clark and Mayer 2011).

Various design principles have been derived according to cognitive load theory (Chandler and Sweller 1991; Merriënboer and Sweller 2005; Paas, Renkl, and Sweller 2003; for an overview see Mayer 2009). One such principle is that online-based learning environments should be designed to reduce cognitive load (CL) and encourage learners to use their free cognitive resources to process essential information (Mayer and Moreno 2003; Merriënboer, Kirschner, and Kester 2003 Merriënboer and Sweller 2005). A possible approach to reduce CL is increasing the interest in the learning material: higher interest promotes the effective use of available cognitive resources (Harp and Mayer 1998; Hidi, Renninger, and Krapp 1992). For example, interest in learning may be enhanced by using a personalized or polite language style (Ginns, Martin, and Marsh 2013; Reichelt, Kämmerer, Niegemann, and Zander, 2014; Stiller and Jedlicka 2010), including quizzing elements, or using new technologies and surprising instructions (e.g., Keller 2010). Both, the motivation-oriented *ARCS* model of Keller (2010) and the *Cognitive-Affective Theory of Learning with Media* by Moreno and Mayer (2007) stress the importance of motivational factors in learning processes in all areas of education—schools, universities, and continuing education. Simply put, motivation stimulates cognitive processing. If learners are interested in an activity, then they will use their available cognitive resources with higher probability to process the information more deeply (e.g., Mayer and Moreno 2003; Merriënboer and Sweller 2005). Consequently, it can be assumed that motivational factors indirectly affect learning, because those factors will increase or decrease learners' cognitive engagement (see also Pintrich 2003).

Research in human information processing, for instance, suggests that learners deal more actively with learning material when they perceive interactions with a partner rather than merely receiving information (Mayer, Fennell, Farmer, and Campbell 2004; Moreno and Mayer 2000, 2004; Reeves and Nass 1996). This assumption is called the interaction hypothesis (Moreno and Mayer 2000). One way to enhance learners' interactions with relevant information is using game elements beides "classical" forms of knowledge transfer. We can assume that the usage of game elements increases the appeal of the learning content. As a limitation, educational design must bear in mind that knowledge acquisition and entertainment factors should be in a highly balanced relationship (Mayer 2016; Plass et al. 2015). For this reason, it is necessary to examine both perspectives: KRS as game and as learning aid.

## 3   KRS: The Object under Investigation

### 3.1   The KRS App

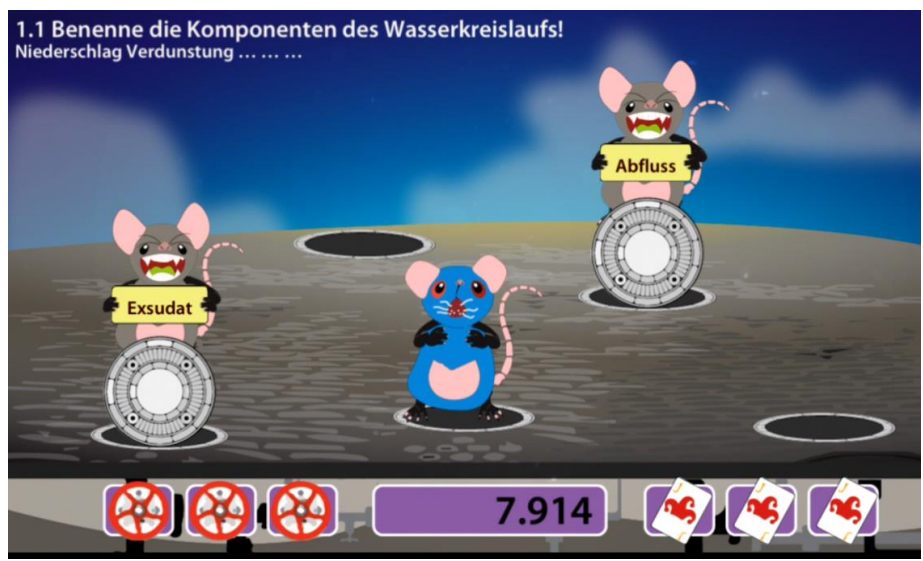

**Fig. 1.** Screenshot of KRS app

The narrative goal of this app is to prevent the sewer rats (German: *Kanalratten*) from achieving world dominance. The learner does this by using subject knowledge to collect nutrients and hunt sewer rats, proven by correctly answering discipline-specific questions (see Fig. 1; Table 1 shows the currently-played sample question). The app has been created in cooperation with Berlin-based company Lernfreak UG, using their LernShooter technology (Buttlar, Kurkowski, Schmidt, and Pannicke 2012). While Lernfreak undertook the technical implementation, we developed the story, graphics, sound and the educational content.

Table 1. Sample question in Fig. 1[1]

| What are the elements of the hydrological cycle? | |
|---|---|
| **Correct** | **Incorrect** |
| Precipitation | Anticipation |
| Evaporation | Rain Cloud |
| Runoff | Astringency |
| Infiltration | Exudation |
| Transpiration | Inspiration |

The app presents multiple-response questions (i.e. questions with multiple correct answers and an accordingly large number of distractors) to the player. As shown in Fig. 1, the player must tap the correct answers, which are held by rats, and omit the incorrect answers. The faster the player hits the correct answers, the larger is the reward. Questions are packaged in so-called levels. During each level, the players are allowed to fail twice, i.e. they have three "lives". Furthermore, they can use three jokers. A joker reveals the correct answer to the current question. The app is connected to the *Apple Game Center*, which registers each player's high score at each level, allowing players to compete with each other. The app is backed by a web-based content management system (CMS), which allows the entry and administration of questions and their instant release to the learners.

### 3.2 Study Aims

Researching game-based learning applications is considered particularly innovative and has gained greatly in popularity (e.g., Karakus et al. 2008; Orvis, Horn, and Belanich 2008; Spector and Ross 2008). Kickmeier-Rust et al. (2006) describe this as a "hot topic," and the discussion continues on how games can benefit learning in educational contexts. As already mentioned, game use has been explored mainly in school settings. A similar problem can be found in instructional design studies (e.g., Mayer 2009). Though previous studies included college students in addition to school students, ID recommendations were mainly tested with college students in psychology courses (e.g., Ginns et al. 2013). Our study addresses this research gap by taking a practical perspective in an educational context with a different target group. In the present explorative study, we examine a game-based learning application with college students in engineering courses. An appropriate design is required to affect learners' motivation positively (e.g., Mayer 2009). Therefore, it is necessary to examine the "balance" between instructional design of learning content and playful design. The explorative study investigates two questions: (1) how do students perceive a game in university teaching, and (2) how far does the KRS impact motivation of students in engineering fields?

---

[1] The original question is in German.

## 4 Study

### 4.1 Overview

The app has been introduced as an innovative, accompanying learning aid to the course of undergraduate education in the area of urban wastewater management, an engineering discipline. It has been intended to be used on the students' smartphones. As it is iOS-based, its target group is restricted by this requirement: only a minor fraction of students owns an iOS device. At the time of a pre-study (August 2013) approximately 60% of students owned a smartphone, with only 20% of those being iOS devices (iPhone and iPad) (Söbke et al., 2014). The KRS app was available to only a part of the course as a voluntary learning aid. An evaluation design investigating the reception and didactic design of KRS was developed according to this fact. It is illustrated in Figure 2. Each row of this figured is presented in a separate section below.

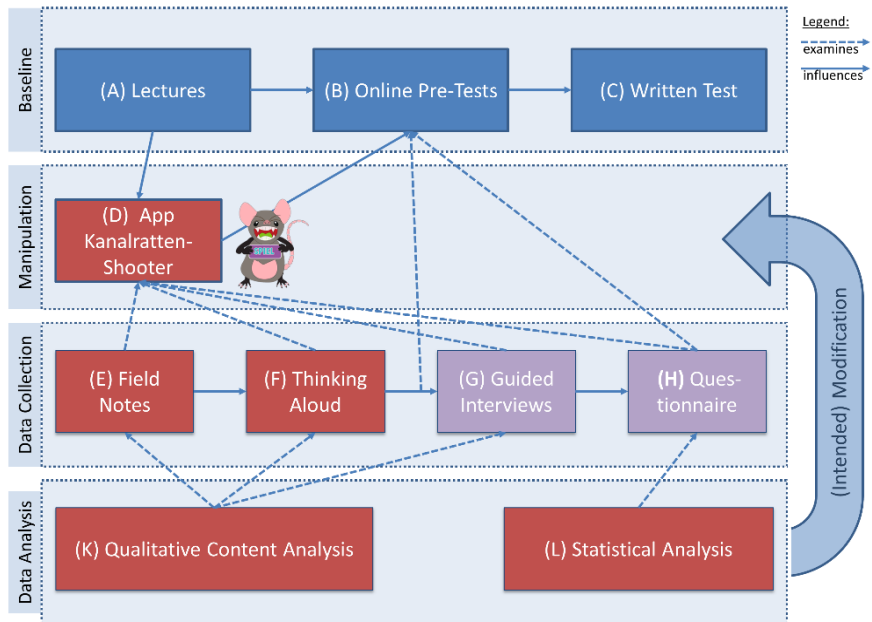

**Fig. 2.** Evaluation Design

### 4.2 Row *Baseline*

Students had created the content for the KRS: multiple response questions that cover the chosen course material (Söbke et al., 2014). The difficulty level of these 160 questions, categorized in 21 topical groups (so-called levels), can be considered as moderate. We integrated these questions into the mandatory assessment schedule in two stages. First, we introduced regular, lecture-accompanying Online Pre-Tests (B). During the semester a total of nine pre-tests were given, each consisting of five lecture-specific questions selected from the KRS corpus. To be admitted to the final test, students had to reach a score of at least 60% in seven of these nine tests. Altogether 97 students registered for the pre-tests. The second stage was the final Written Test (C). A first 30-minute section of this written test was reserved for questions previously contained in the online pre-tests. In contrast to the pre-tests, no answers and distracters were offered. The rest of the exam was made up by calculation tasks. The final test was taken by 70 students.

### 4.3 Row *Manipulation*

The first lecture introduced students to the assessment structure, presenting the KRS app as a learning tool for those students who owned an iOS device. The number of students who had access to the KRS app was counted via a questionnaire during the registration for the pre-tests as 17.

### 4.4 Row *Data Collection*

The multi-method design started with *Field Notes* (E) or game diaries. Five persons without content-affinity operated the KRS app at least 5 minutes a day for 2 weeks and took notes on their observations and experiences (Kießig, Köchy, Lang, Pohl, and Schilling 2015). The findings of these field notes influenced the next step, *Thinking Aloud* (F) combined with a guided interview (Frommann 2005). We observed and recorded a total of 16 thinking aloud sessions in mid-semester consisting of 5 minutes of game play and a subsequent interview. All the participants were registered in the course, so they showed content-affinity. Six of them had prior KRS app experience. Students were instructed to play a level of their choice and to explain their thoughts aloud as they occurred. Finally, we asked about their experience with the app, guiding the interview with a set of questions. Altogether, each session lasted for about 25 minutes. Thereafter the recordings were transcribed and coded, i.e. assigned to summarizing statements. In the final lecture we conducted a Guided Interview (G) (n=30). Participants of this interview have been both KRS users and non-KRS users. Subsequently, an online Questionnaire (H) was issued (n=16), which could be answered before or after the Written Test (C).

### 4.5 Row *Data Analysis*

An essential part of the evaluation is the *Qualitative Content Analysis* (K). It is based on the *Field Notes*, the *Thinking Aloud* sessions, and the *Guided Interviews*. The *Statistical Analysis* (L) processed the results of the *Questionnaire* (H).

## 5 Results

### 5.1 Field Notes

Field notes were taken by persons who were originally not interested in the content, but in evaluation of the app's efficacy. This setting led to insights—some harshly-expressed—purely about the game and its mechanics, and less influenced by the content. Players generally reported low intrinsic motivation, as demonstrated by such comments as "In reality I would not play this app" or "This app provides no gaming fun at all" [2]. The missing intrinsic motivation was also indicated by playing times, which had to be supplied in the field notes: in most cases the five-minute requirement was met exactly. Field notes also contained comments about possible improvements and missing features. One criticism was that answer signs are displayed too briefly in cases where answer texts are very long, so a text-length dependent display time would help. A few probands pointed out the lack of competition between fellow players[3]. Players did not consider *Apple Game Center*-based rankings sufficient as they interfered with another characteristic marked as deficit: players complained about the lack of transparency of the number of reached points. Currently points are awarded almost continuously (each millisecond counts) according to the time used to hit each correct answer.

In terms of didactical design, players recognized that the first level comprises only five simple questions, and they complained about the steep increase in difficulty as the levels advanced and grew in size up to eleven questions. Another recurring statement was that prior knowledge was required to solve the tasks. This could suggest that the didactical design needs to be improved: learning is not integrated seamlessly, but players are pointed to their deficits too harshly. Content design was mentioned also: in some cases, the answers to a question form a complete sentence and therefore must be tapped in a defined order (see Table 2). This characteristic has been criticized as arbitrary and not intuitive.

**Table 2.** Example of sentence-based question

| What is a water network? | |
|---|---|
| **Correct** | **Incorrect** |
| Transportation system … | The labyrinth … |
| … for water … | … consisting of pipes … |
| … using … | … for sewage water transport … |
| … creeks and … | … in terms of a … |
| … rivers. | … a sewer system. |

Probands were aware that learning is one goal of this app. Thus learning was mentioned in the field notes. Some statements acknowledged generally that learning occurs during gaming: "The drill & exercise methods seems to be successful, because

---

[2] The original comments are in German; translation is by the authors.
[3] In accordance with these findings we could identify competition between friends as an important characteristic of commercial quiz apps (Söbke, 2015).

I can remember parts of the content." Some comments drew the connection between fun and learning: "it becomes boring very fast and therefore there is no learning success" or "the game still is fun, because there is a learning effect." Other statements pointed to the difficulty of questions that reduced the success in the game. Increased difficulty was attributed to the number of answers (e.g., eight answers were perceived as difficult.). However, success seems to be a foundation for fun. Other comments point to potential "dysfunctional" learning processes. For instance, one person stated that it was easier to remember the wrong answers and to avoid them. Another proband declared that she recognized the correct answers according to their visual form (i.e. the length of the answer and the type face), but not semantically. Additionally, sentence-based answers were selected according to their syntax (e.g., ellipsis points, punctuation marks).

### 5.2 Thinking Aloud

Table 3 shows statements that were made by at least half of the KRS-experienced probands during the interviews. Column 3 shows the number of probands who affirmed the statement explicitly. The rest of the probands either denied the statement or did not mention it. The interview was not standardized, but only guided. The researchers assigned the statements to categories.

**Table 3.** Thinking Aloud: coded statements of the previous-knowledge group, given by at least half of the group (n=6)

| Category | Statement | Number of probands |
|---|---|---|
| Content | Sentence-based answers are critical | 5 |
|  | Enumerations (Example in Table 1) preferred (in contrast to sentences, example in Table 2) | 3 |
|  | Number of questions in a level is appropriate | 3 |
| Game mechanics | Speed of flashing answers is ok | 5 |
|  | Incomprehensible algorithm for points is irritating | 4 |
|  | Jokers have been used | 4 |
|  | No Apple Game Center account | 3 |
| Hardware | Played on iPhone | 4 |
| Learning | Impression of learning, when using the app | 3 |
| Previous knowledge | Experience with other games for learning | 4 |
| Usability | Delays during loading process irritate | 3 |
| Usage profile | Have not seen the tutorial | 4 |

| | Facilitation as dictionary | 4 |
| --- | --- | --- |
| User acceptance | For course members a good choice | 6 |
| | Would recommend it to course members | 4 |
| | App enriches the daily learning routine | 4 |
| | App supports regular learning content related engagement | 3 |
| | App is not the primary reason for learning content related engagement | 3 |
| | Online pre-tests spur usage of the app | 3 |

The number of questions per level having been marked in the field notes as too large is considered appropriate. The algorithm for calculating the number points a player is rewarded with, is affirmed as not understandable, too. Because half of the players did not use an *Apple Game Center* account to deliver their results to a ranking list, they voluntarily missed the chance to compete with their course mates. Most respondents played the app on an iPhone, which is relevant as its screen size is smaller than the one of an iPad. Although only half of the players said they felt that they were learning and valued the app as a positive way to engage with the course content, more than half expressed a positive attitude towards the app. They would recommend it to fellow course members and value it as an enrichment of their learning routine. Further, they stated that it helps them to deal more regularly with learning content. However, for most of the players the app is just a means to reach the course goals: They used it as a dictionary in order to accomplish the online-pre-tests, and they stated that the app is not primarily a reason to engage in the course content.

The second *Thinking Aloud* experiment was conducted by the group who recorded the field notes. They interviewed 10 course members who had no prior app knowledge. The interviewers had access to the questions of the guided interview, but they were not aware of the results of the first Thinking Aloud experiment. Table 4 again shows statements that were made consistently by at least half of the interviewees. Some of the statements of the first experiment are affirmed: the level size was considered appropriate, the students had the impression of learning, they would recommend the app to their course mates, and they agreed that the app would enrich the learning routine. However, there was one contradicting statement: the flashing speed, in which the answers appear and disappear, is considered too fast—a characteristic which could easily be changed in an updated version of the KRS app. Summing up, the findings of the second experiment confirm that the students received the app positively: They had a good overall impression of the app, felt it had an appealing design, and considered it as an appropriate tool for learning.

**Table 4.** Thinking Aloud: coded statements of the no-previous-knowledge group, given by at least half of the group (n=10); green: confirming, red: contradicting (see Table 3)

| Category | Statement | Number of probands |
| --- | --- | --- |

| Content | Number of question in a level is appropriate | 8 |
|---|---|---|
| **Game mechanics** | Speed of flashing answers is not ok | 6 |
| Learning | Recognition of the course content in the app | 9 |
| | Impression of learning, when using the app | 7 |
| Previous knowledge | My media/computer literacy is good | 5 |
| Usability | All questions was readable | 8 |
| User acceptance | I would use this app for learning | 9 |
| | In general learning with apps makes sense | 9 |
| | The app design is appealing | 8 |
| | The app would enrich the daily learning routine | 8 |
| | Overall-impression of the app is good | 7 |
| | I would recommend it to course members | 7 |
| | It was fun to see the content of the course in the app | 7 |

In addition to these categorized and summarized interview results, respondents made further noteworthy remarks. In general, they felt that using the app contributed positively to an innovative image of the university. One student reported that she uses the KRS app during television ad breaks. As her smartphone is available in front of the television, the app is accessible, and accessibility leads to learning. She mentioned that this behavior soothes her conscience to have learned for the final exam. Another partly surprising result was that players did not consider game elements important; reaching a new high score was not rated as a desirable goal of the game. This may be attributable to the non-transparent reward schedule or to the additionally required *Apple Game Center* account.

Several strategies were used to create false answers, e.g. words with a spelling error or words having a funny meaning in the context of the question (Haladyna and Rodriguez 2013; Söbke et al., 2014). The interviews revealed that all these strategies have their supporters and their opponents. The only exception is the sentence-based approach (which has been used to define the correct answers, but which also applies to false answers).

Furthermore, the interviews delivered suggestions for improving the app's usability. For instance, on the iPhone level selection does not indicate that there are more than three levels. Additionally, it is hard to distinguish between a "killed" and an "escaped" rat, therefore these graphics have to be reworked.

### 5.3 Guided Interview

During the last lecture we conducted guided interviews with 30 students (KRS players and non-KRS players). One part of this interview referred to three sample questions, previously distributed via the KRS app and online pre-tests. Students were asked to write down the answers to the first two questions and to estimate the completeness (i.e. the percentage of correct answer elements) of their answer. For the first question, the students wrote down on average 63% of the answers. They estimated 46% of completeness. For the second question only 19% of correct answers were given, though students estimated 40%. The poor performance on the second question probably has two main causes: The question partly overlapped with another question, so some students confused the answers, and the question was complex as it had seven correct answer parts.

For the third question we provided students with all correct and incorrect answers and asked them to write down the question to which these answers applied. Only one student of 30 was able to do so. This result seems to contradict one of the claims of the field notes, that players are learning the false answers.

In general, the results of these interviews contributed to the content of the final questionnaire. For example, some students expressed doubt that the app and online pre-tests were sufficient for final test preparation. This led us to integrate questions from the Questionnaire on Current Motivation (QCM) (Rheinberg, Vollmeyer, and Burns 2001).

### 5.4 Questionnaire

The general aim of the questionnaire was to gain further insights into the user behavior, for example the perceived app´s attractiveness, task difficulty, or usage contexts of the KRS. Overall 16 of 57 students played the KRS; 12 students had a private device, and four played on a device loaned to them. Because so many students did not have a private device, one practical implication of the study is that access to a device should be ensured. One positive sign of the app's attractiveness is that 25% of the players borrowed a device so that they could use the app. Only three students played regularly—at least three times a week—during the semester, and they played in short sessions. On average, a game session lasted 5 to 10 minutes, which corresponds to 30 seconds per question. In addition, students did not play the KRS intensively. The KRS has 20 levels and every student played each level twice. As reason for only temporary use of the KRS students stated not to have "enough time to learn" and "learning for the exam at the end had priority." The questionnaire results indicate that the KRS was not suitable for learning during the examination phase, because this time was very stressful and "crammed" with tasks. 12 out of 16 students who used the KRS reported that they mainly used their written lecture notes instead of the KRS during the examination phase. Consequently, we hypothesize that students will not play in stressful learning contexts. Only two students reported that they used the KRS during the examination phase as exam preparation and they positively noted the combination of relaxation and learning through the game elements of the KRS app. This leads to the question, in which part of the learning phase the KRS is helpful and supportive? The results provide evidence that the game elements are inappropriate under time pressure, such as just before an exam.

To gain more insight into the use of the KRS, we included questions about several occasions of usage. Thirteen of sixteen students (81%) reported that they mainly used the KRS as a "reference book". They played only to figure out the correct answers. Thereafter, they used their notes for further learning activities. The KRS was used selectively by 31% of the students to learn the content of single lectures. Here, the focus was not on the game elements of the KRS but rather on knowledge acquisition. Only four students used the KRS every day during "commercial breaks" or while "waiting for the train."

Further, the students were asked for their primary motivational elements of playing the KRS app. The results showed that the players were pleased most when they had properly managed a level (10 of 16 respondents selected this item). Nine respondents were motivated when they answered a question correctly. It was also motivating when a new, unknown level was played successfully for the first time (6 of 16 respondents). In summary, respondents were particularly delighted with their own learning performance. The game elements themselves were not perceived as a motivational incentive. However, five students acknowledged witty distractors, but achieving "high scores" was more important for them.

To examine the motivational aspects in more detail, we included items from each dimension of the Questionnaire on Current Motivation (QCM). The following table shows the selected items and the median. The scale ranges from *1 (not applicable)* to *7 (applicable)*.

**Table 5.** QCM: Results

| Dimension | Items | M | SD |
| --- | --- | --- | --- |
| Confidence of success | I think everybody can learn the correct answers with the KRS. | 5.0 | 1.507 |
| | Probably, I will not learn the questions adequately fairly with the KRS. | 4.0 | 1.342 |
| Fear of failure | I am a bit concerned when I think of playing the KRS. | 2.4 | 2.012 |
| | I feel under pressure to reach good results in the KRS. | 1.9 | 1.698 |
| Challenge | I make a great effort when I play the KRS. | 2.8 | 1.503 |
| | Playing the KRS is a challenge for me. | 3.1 | 1.340 |
| Interest | Playing the KRS is fun. It is a welcome diversion. | 4.1 | 1.354 |
| | I would play the KRS in my spare time, too. | 2.6 | 1.521 |

Although students estimated the probability of success as being very high, they doubted that the KRS is sufficient for learning. This result indicates that other learning materials such as lecture notes will be necessary to complete the learning process and increase learners' satisfaction. As expected, the fear of failure is low while playing the app because there are no external penalties or incentives. The results in the dimension *Challenge* were similar. The students' willingness to use the KRS in their spare time was rather low on average. However, the KRS was considered to support the learning process. To examine the current motivation, we included the item "I'm more motivated to start the KRS and play a level than to recapitulate the lecture scripts". The average value was 3.5 (SD=2.034). Students (13 of 16) recommended that the KRS should be part of future lectures, too.

Previous research has shown a relationship between motivation and perceived task difficulty (e.g., Paas 1992; Sweller, Merriënboer, and Paas 1998). Therefore, we included the item "The arithmetic problems in the lectures are a bit more difficult than the questions in the KRS." The results for the perceived task difficulty showed an average of M=5.6 (SD=1.141), which acknowledges the item.

By an open text field, we asked for supplementary comments and impressions about the KRS app. Remarkable responses are summarized in Table 6.

**Table 6.** Students' perceptions

| Statements (short version) |
|---|
| "It's a funny way to remember the content." |
| "It's a good idea, but not all students have a suitable device." |
| "Some students were able to use the game when they lent a device from friends; however, they could not play regularly. Therefore, the positive effect on learning outcome has failed." |
| "I have only played to gather the answers of test questions as quickly as possible every week." |
| "It's a nice diversion." |
| "It helps to retain questions and their answers better." |

## 6 Limitations and Discussion

This explorative study investigated the effect of a game-based learning app on learners' motivation and analyzed the perceived impact and appeal as well as the role of game elements as a potential motivational incentive.

In general, the findings of game diaries, think aloud methods, and semi-structured interviews imply that the subjective leaners' perception of the KRS app depends on various influencing factors. Based on the analysis of game diaries, the joy of playing is not sufficient to encourage learners' intrinsic motivation. The findings of the qualitative interview analysis (Mayring 2007) assumes that learners often used the KRS app in order to calm their consciences ("Today I prepared myself for the exam at least partly."). Future research should be conducted to determine if this calming function has negative effects on students' learning efforts in general and could therefore lead to even worse results. Specifically, the flexible use (e.g., usage of short time gaps) and the low physical barriers (e.g., "spontaneous learning on the sofa") motivated some students to play the app. Furthermore, a lower perceived task difficulty was described in comparison to the use of the lecture notes. However, the assumed importance of game elements was not confirmed through learners' statements. Especially, the app's usage was not considered helpful shortly before the exam. An assumption that the KRS app could serve as an aid for mental warm-up before lectures was not confirmed. Participants indicated that the KRS app partly reinforced collaborative learning. In addition, the interview analysis and the data of thinking aloud showed several recommendations for continuation as well as modification of the app design: For example, the background music was rated as positive ("The music was very good"), but respondents recommended that the evaluation scheme should be made more transparent and display times of answers should be adjusted.

Content creation is a crucial task. Completing sentences has been identified as not matching the design of the KRS. In addition, the level design requires more effort in order to harmonize the complexity of the different levels. As a further improvement we suggest focusing on content-creation in order to generate questions that activate learners' prior knowledge. Currently students state that they are not inspired to reflect on questions.

We received one remarkable contribution during the guided interview: "Studying is a voluntary, self-determined process. A person, who is not interested in a certain topic, will not be guided to efficient learning by such a compulsory measure. Motivation has to be intrinsic." Although this is an isolated remark, it points to how using an app may conflict with the self-conception of graduate education, where freedom and personal responsibility of the individual are important values. Nevertheless, this argument does not question the app itself, and other comments welcomed the app's enforcement of learning activities.

Although the educational effects seem unclear, students' almost unanimous recommendation was to keep the KRS app as an educational element of the course. They indicated that the app contributed to their positive attitude toward the course as up-to-date education and it has been a "witty means of learning". From a perspective of Instructional Design, the embedding of the KRS app into the course has to be improved so that its use becomes more natural.

## 7  Conclusions and Outlook

The conducted study was designed as an explorative investigation and combined the field of Instructional Design with Digital Game-Based Learning. We used a multi-method-strategy to investigate learners' perception of a mobile learning app and its impact on their motivation. Certain phenomena could be recognized using multiple methods, e.g. from the perspective of game design, the unfortunately low motivational impact. A positive effect is the temporally and spatially flexible use of the app, which revealed new occasions of learning. However, still unclearand therefore a subject for future researchis the app's effectiveness and any probable negative effects on learning.

One remarkable finding is that game elements seem to contribute to learners' motivation only at a low level. This observation may be ascribable in parts to characteristics of the target group of grown-up and supposed to act rationallyengineering students. The lack of motivational effects requires further work on game design in order to provide an app with a greater impact on intrinsic motivation. As a more feasible alternative in the short run, we suggest investigating how non-game apps perform in the given context. This might be specifically of interest, because game apps are considered as being ineffective in the stage of immediate exam preparation. In general, this finding points to a missed result of the experiment: ensuring steady learning during the course and not right before an exam. Students' high regard of the KRS provision in an educational contextdespite of all its identified weaknesseswas another remarkable result of this study. Currently, such an app seems to benefit from its uniqueness in formal education. However, on the long term it should be worked on its deficiencies.

From the view of serious game design the study led to a number of potential detailed design changes to increase the app's capacity to spur intrinsic motivation and engagement. The introduction of a list of friends, groups, matches against friends, and rankings are considered as important elements. Additionally, a more transparent reward schedule, a random sequence of questions within a level, more detailed statistical information, and unlocked levels as rewards belong to required changes.

Furthermore, the results of the investigation contributed to hypotheses. The following questions can be starting points for future work: How do game-based learning apps affect learning when the high score is relevant and the performance orientation is focused? How does the KRS app influence learners' mental effort, and what effect does this have on the learning outcome? How could a formal learning context be designed to integrate the app more seamlessly and promote course-accompanying learning?

Finally, game-based learning provides no silver bullet to learning processes. However, it is an additional "weapon" in the instructional arsenal. The study provides an example for the further development of game-based learning in alignment with methodologies of Instructional Design. The received concordant positive acknowledgement of students—even in a case with unclear effectiveness—can be valued as encouragement to further pursue such attempts.

## 8   References

Bartle, R. A. (1996). Hearts, clubs, diamonds, spades: Players who suit MUDs. Journal of MUD Research, 1(1), 19.

Bruckman, A. (1999). Can educational be fun? In *Game Developer's Conference*, San Jose, California (pp. 75–79).

Buttlar, R. von, Kurkowski, S., Schmidt, F. A., & Pannicke, D. (2012). Die Jagd nach dem Katzenkönig. In W. Kaminski & M. Lorber (Eds.), *Gamebased Learning: Clash of Realities 2012* (pp. 201–214). München: Kopäd..

Chandler, P., & Sweller, J. (1991). Cognitive load theory and the format of instruction. *Cognition and Instruction*, 8(4), 293–332.

Clark, R. C., & Mayer, R. E. (2011). E-Learning and the Science of Instruction: Proven Guidelines for Consumers and Designers of Multimedia Learning (3rd ed.). San Francisco: Pfeiffer.

Csikszentmihalyi, M. (1990). *Flow: The Psychology of Optimal Experience*. New York: Harper and Row.

Deterding, S., Dixon, D., Khaled, R., & Nacke, L. (2011). From game design elements to gamefulness: Defining gamification. In *Proceedings of the 15th International Academic MindTrek Conference: Envisioning Future Media Environments* (pp. 9–15). New York: ACM.

Djaouti, D., Alvarez, J., & Jessel, J.-P. (2011). In P. Felicia (Ed.), *Handbook of Research on Improving Learning and Motivation through Educational Games: Multidisciplinary Approaches* (pp. 118–136). Hershey, PA, USA: IGI Global.

Ebner, M., & Holzinger, A. (2007). Successful implementation of user-centered game based learning in higher education: An example from civil engineering. *Computers and Education*, *49*(3), 873–890.

Egenfeldt-Nielsen, S. (2005). Beyond edutainment: Exploring the educational potential of computer games. Raleigh, North Carolina: lulu.com..

Frommann, U. (2005). Die Methode "Lautes Denken." Retrieved October 5, 2014, from https://www.e-teaching.org/didaktik/qualitaet/usability/Lautes Denken_e-teaching_org.pdf

Garris, R., Ahlers, R., & Driskell, J. E. (2002). Games, motivation, and learning: A research and practice model. *Simulation Gaming, 33*(4), 441–467.

Gee, J. P. (2004). Learning by design: Games as learning machines. *Interactive Educational Multimedia*, (8), 15–23.

Gee, J. P. (2008). What Video Games Have to Teach Us about Learning and Literacy. New York: Palgrave Macmillan.

Ginns, P., Martin, A. J., & Marsh, H. W. (2013). Designing instructional text in a conversational style: A meta-analysis. *Educational Psychology Review*, *25*(4), 445–472.

Habgood, M. P. J., & Ainsworth, S. E. (2011). Motivating children to learn effectively: Exploring the value of intrinsic integration in educational games. *Journal of the Learning Sciences*, *20*(2), 169–206.

Haladyna, T. M., & Rodriguez, M. C. (2013). *Developing and Validating Test Items*. New York: Routledge..

Harp, S. F., & Mayer, R. E. (1998). How seductive details do their damage: A theory of cognitive interest in science learning. *Journal of Educational Psychology*, *90*(3), 414–434.

Hidi, S., Renninger, K. A., & Krapp, A. (1992). The present state of interest research. In K. A. Renninger, S. Hidi, & A. Krapp (Eds.), *The Role of Interest in Learning and Development* (pp. 433–446). New York: Psychology Press.

Kapp, K. M. (2012). The Gamification of Learning and Instruction: Game-based Methods and Strategies for Training and Education (1st editio). San Francisco: Pfeiffer.

Karakus, T., Inal, Y., & Cagiltay, K. (2008). A descriptive study of Turkish high school students' game-playing characteristics and their considerations concerning the effects of games. *Computers in Human Behavior*, *24*(6), 2520–2529.

Keller, J. M. (2010). Motivational Design for Learning and Performance: The ARCS Model Approach. New York: Springer.

Kickmeier-Rust, M., Schwarz, D., Albert, D., Verpoorten, D., Castaigne, J.-L., & Bopp, M. (2006). The ELEKTRA project: Towards a new learning experience. M3 - Interdisciplinary Aspects on Digital Media & Education. *Proceedings of the 2nd Symposium of the WG HCI & UE of the Austrian Computer Society*, *3*(4), 19–48.

Kießig, F., Köchy, J., Lang, Y., Pohl, N., & Schilling, F. (2015). *NASS und die Effektivität des KanalrattenShooters*. Bauhaus-Universität Weimar.

Konert, J., Göbel, S., & Steinmetz, R. (2013). Player and learner models: Independency of Bartle, Kolb and BFI-K (Big5). In P. Escudeiro & C. Vaz de Carvalho (Eds.), *Proceedings of the 7th European Conference on Games Based Learning* (pp. 329–335).

Malone, T. W., & Lepper, M. R. (1987). Making learning fun: A taxonomy of intrinsic motivations for learning. *Aptitude, Learning, and Instruction*, *3*, 223–253.

Mayer, R. E. (2009). *Multimedia Learning* (2nd ed.). New York: Cambridge University Press.

Mayer, R. E. (2016). What should be the role of computer games in education? *Policy Insights from the Behavioral and Brain Sciences.* http://doi.org/10.1177/2372732215621311

Mayer, R. E., Fennell, S., Farmer, L., & Campbell, J. (2004). A personalization effect in multimedia learning: Students learn better when words are in conversational style rather than formal style. *Journal of Educational Psychology*, *96*(2), 389–395.

Mayer, R. E., & Moreno, R. (2003). Nine ways to reduce cognitive load in multimedia learning. *Journal of Educational Psychology*, *38*(1), 43–52.

Mayring, P. (2007). On generalization in qualitatively oriented research. *Forum: Qualitative Social Research*, *8*(3), Art. 26, http://nbn–resolving.de/urn:nbn:de:0114–f qs0703262.

Meier, C., & Seufert, S. (2003). Game-based learning: Erfahrungen mit und Perspektiven für digitale Lernspiele in der beruflichen Bildung. In A. Hohenstein & K. Wilbers (Eds.), *Handbuch E-Learning*. Köln: Fachverlag Deutscher Wirtschaftsdienst.

Merriënboer, J. J. G. van, Kirschner, P. A., & Kester, L. (2003). Taking the Load Off a Learner's Mind: Instructional Design for Complex Learning. *Educational Psychologist*, *38*(1), 5–13.

Merriënboer, J. J. G. van, & Sweller, J. (2005). Cognitive load theory and complex learning: Recent developments and future directions. *Educational Psychology Review, 17*(2), 147–177.

Michael, D. R., & Chen, S. L. (2005). *Serious Games: Games That Educate, Train, and Inform.* Mason, OH, USA: Course Technology.

Moreno, R., & Mayer, R. E. (2000). Engaging students in active learning: The case for personalized multimedia messages. *Journal of Educational Psychology*, *92*(4), 724.

Moreno, R., & Mayer, R. E. (2004). Personalized messages that promote science learning in virtual environments. *Journal of Educational Psychology, 96*(1), 165.

Moreno, R., & Mayer, R. E. (2007). Interactive multimodal learning environments. *Educational Psychology Review*, *19*(3), 309–326.

Orvis, K. A., Horn, D. B., & Belanich, J. (2008). The roles of task difficulty and prior videogame experience on performance and motivation in instructional videogames. *Computers in Human Behavior, 24*(5), 2415–2433.

Paas, F. G. W. C. (1992). Training strategies for attaining transfer of problem-solving skill in statistics: A cognitive-load approach. *Journal of Educational Psychology, 84*(4), 429–434.

Paas, F. G. W. C., Renkl, A., & Sweller, J. (2003). Cognitive load theory and instructional design: Recent developments. *Educational Psychologist*, *38*(1), 1–4.

Papert, S. (1998). Does easy do it? Children, games, and learning. *Game Developer*, *5*(6).

Pintrich, P. R. (2003). A motivational science perspective on the role of student motivation in learning and teaching contexts. *Journal of Educational Psychology*, *95*(4), 667–686.

Plass, J. L., Homer, B. D., & Kinzer, C. K. (2015). Foundations of game-based learning. *Educational Psychologist*, *50*(4), 258–283.

Ratan, R., & Ritterfeld, U. (2009). Classifying serious games. In U. Ritterfeld, M. Cody, & P. Vorderer (Eds.), *Serious games: Mechanisms and effects* (pp. 10–22). New York: Routledge..

Reeves, B., & Nass, C. (1996). *The Media Equation: How People Treat Computers, Television, and New Media Like Real People and Places*. Stanford: The Center for the Study of Language and Information Publications.

Reichelt, M., Kämmerer, F., Niegemann, H. M., & Zander, S. (2014). Talk to me personally: Personalization of language style in computer-based learning. Computers in Human Behavior, 35, 199–210.

Rheinberg, F., Vollmeyer, R., & Burns, B. D. (2001). QCM : A questionnaire to assess current motivation in learning situations. *Diagnostica*, *47*, 57–66.

Rockwell, G. M., & Kee, K. (2011). Game studies: The leisure of serious games: A dialogue. *Game Studies - the International Journal of Computer Game Research*, *11*(2). Retrieved from http://gamestudies.org/1102/articles/geoffrey_rockwell_kevin_kee

Roppelt, B. (2014). *Gamification in der Hochschullehre durch eine Quiz-App*. Universität Passau. Retrieved from http://www.eislab.fim.uni-passau.de/files/publications/2014/roppelt2014uniquiz.pdf

Ryan, R., & Deci, E. (2000). Intrinsic and extrinsic motivations: Classic definitions and new directions. *Contemporary Educational Psychology*, *25*(1), 54–67.

Simões, J., Redondo, R. D., & Vilas, A. F. (2013). A social gamification framework for a K-6 learning platform. *Computers in Human Behavior* 29(2), 345–353.

Söbke, H. (2015). Space for seriousness? Player Behavior and Motivation in Quiz Apps. In K. Chorianopoulos & et al. (Eds.), Entertainment Computing – ICEC 2015 14th International Conference, ICEC 2015 Trondheim, Norway, September 29 – October 2, 2015 Proceedings (pp. 482--489). Cham: Springer. http://doi.org/10.1007/978-3-319-24589-8_44

Söbke, H., Bröker, T., & Kornadt, O. (2013). Using the master copy - Adding educational content to commercial video games. In 7th European Conference on Games Based Learning, ECGBL 2013 (Vol. 2).

Söbke, H., Chan, E., Buttlar, R. von, Große-Wortmann, J., & Londong, J. (2014). Cat King's Metamorphosis - The Reuse of an Educational Game in a Further Technical Domain. In S. Göbel & J. Wiemeyer (Eds.), Games for Training, Education, Health and Sports (Vol. 8395, pp. 12–22). Darmstadt: Springer International Publishing. http://doi.org/10.1007/978-3-319-05972-3_3

Spector, J. M., & Ross, S. M. (2008). Special thematic issue on game-based learning. *Educational Technology Research and Development*, *56*(5–6), 509–510.

Squire, K. R. (2011). Video Games and Learning: Teaching and Participatory Culture in the Digital Age. New York: Teachers College Press.

Stiller, K. D., & Jedlicka, R. (2010). A kind of expertise reversal effect: Personalisation effect can depend on domain-specific prior knowledge. *Australasian Journal of Educational Technology, 26*(1), 133–149.

Sweller, J., Merriënboer, J. J. G. van, & Paas, F. G. W. C. (1998). Cognitive architecture and instructional design. *Educational Psychology Review*, *10*(3), 251–296.

Yee, N. (2006). Motivations for play in online games. Cyberpsychology & Behavior : The Impact of the Internet, Multimedia and Virtual Reality on Behavior and Society, 9(6), 772–775.